\documentclass[11pt,a4paper]{article}
\usepackage[utf8]{inputenc}
\usepackage[T1]{fontenc}
\usepackage{amsmath}
\usepackage{amsfonts}
\usepackage{graphicx}
\usepackage{caption}
\usepackage{booktabs}
\usepackage{multirow}
\usepackage{siunitx}
\usepackage{amssymb}
\usepackage{color}
\usepackage{colortbl}
\usepackage{siunitx}
\usepackage{booktabs}
\usepackage{longtable}
\usepackage[left=2.50cm, right=2.50cm, top=2.50cm, bottom=2.50cm]{geometry}
\usepackage[onehalfspacing]{setspace}
\usepackage{placeins}
\usepackage{authblk}
\usepackage{lipsum}

\usepackage{apacite} 			
\usepackage{natbib}	

\usepackage[title,titletoc,toc,page]{appendix}
\appendixpageoff
\appendixtocoff

\title{Small area prediction of counts under machine learning-type mixed models}
\author[*]{Nicolas Frink}
\author[*]{Timo Schmid}
\affil[*]{\small{Institute of Statistics, Otto-Friedrich-Universit\"{a}t Bamberg, Bamberg, Germany}}
\date{}

\begin{document}

\maketitle
\onehalfspacing
\normalsize
\setlength{\parindent}{12pt}

\begin{abstract}
This paper proposes small area estimation methods that utilize generalized tree-based machine learning techniques to improve the estimation of disaggregated means in small areas using discrete survey data. Specifically, we present two approaches based on random forests: the Generalized Mixed Effects Random Forest (GMERF) and a Mixed Effects Random Forest (MERF), both tailored to address challenges associated with count outcomes, particularly overdispersion. Our analysis reveals that the MERF, which does not assume a Poisson distribution to model the mean behavior of count data, excels in scenarios of severe overdispersion. Conversely, the GMERF performs best under conditions where Poisson distribution assumptions are moderately met. Additionally, we introduce and evaluate three bootstrap methodologies - one parametric and two non-parametric - designed to assess the reliability of point estimators for area-level means. The effectiveness of these methodologies is tested through model-based (and design-based) simulations and applied to a real-world dataset from the state of Guerrero in Mexico, demonstrating their robustness and potential for practical applications.
\end{abstract}


\section{Introduction}\label{sec:1}
Linear mixed models (LMM) are a staple for analyzing unit-level survey data and estimating area-level means. These models account for the hierarchical structure of observations through random effects. An example of such a model is the nested error regression model \citep{Battese1988}, which requires the availability of unit-level survey data and administrative auxiliary information, such as data from a register or census. However, it is crucial to recognize the limitations imposed by the distributional and structural assumptions of these models, which may not always hold in small area estimation (SAE) applications \citep{Molina2015}. For example, when working with LMMs, it is necessary to assume a linear relationship between the covariates and the outcome variable. This assumption may not always align with empirical evidence. Thus, it is critical to ensure the validity of model assumptions for optimal results and predictive performance of model-based SAE. If assumptions are not met, parameter estimates may be biased, and the reliability of mean squared error (MSE) estimates may be compromised \citep{JiangRao2020}.

To circumvent the parametric assumptions inherent in LMMs, employing machine learning techniques is a viable methodological alternative. These techniques are capable of extracting predictive relationships from data, encompassing complex interactions among covariates, without relying on explicit model assumptions \citep{Hastie_etal2009, Varian2014}. \citet{Krennmair2023} provide a framework utilizing tree-based machine learning methods for SAE. This framework presents a non-linear, data-driven, and semi-parametric approach for continuous variables to estimate area-level means using Mixed Effects Random Forests (MERF). Random forests \citep{Breiman2001} are renowned for their robust predictive performance, particularly in the presence of outliers, and they inherently address model-selection issues \citep{Biau2016}. MERFs incorporate these benefits while also modeling hierarchical dependencies. Originally developed for continuous target variables similar to LMMs, \citet{Frink2024} have expanded the application of MERFs to estimate poverty indicators using binary response variables within the context of SAE. However, their focus is limited to binary variables, and the behavior of MERFs with count data remains unexplored.

Various methods for SAE with count data have been investigated. Parametric models are discussed by \citet{Ghosh1998}, \citet{Jiang2006}, \citet{Dreassi2014}, \citet{Chen2015}, \citet{Boubeta2016}, \citet{Hobza2016}, \citet{Boubeta2017} and \citet{Chandra2017}. In contrast, \citet{Chambers2014} and \citet{Tzavidis2015} have examined semi-parametric unit-level models utilizing M-quantiles, which only partially rely on the distributional assumptions of the Poisson and Negative Binomial families. However, all of these methodologies for count data assume a linear relationship within the systematic component of the underlying model.

Therefore, this paper introduces semi-parametric approaches to SAE for count data using random forests, which allow for a flexible specification of the relationship between the outcome variable and the covariates. Building on the work of \citet{Krennmair2023} and \citet{Frink2024}, we extend the (generalized) MERF approaches - originally designed for continuous and binary data - to scenarios involving count outcomes.

The generalized MERF (GMERF) is presented as a flexible and data-driven approach that employs the Poisson distribution to model the mean behavior of count data. Furthermore, a parametric MSE bootstrap scheme is proposed to evaluate the uncertainty associated with area-level estimates. While the Poisson distribution is commonly used to model the mean behavior of count data, it often underestimates the variability in cases of overdispersion. Overdispersion occurs when the observed variance of the response variable exceeds the variance predicted by the Poisson distribution. It is a prevalent issue in count data analysis and critically affects the interpretation of uncertainty measures \citep{VerHoef2007}. To address this challenge and ensure robust uncertainty estimators, even when the Poisson distribution's assumptions are not met, we introduce a non-parametric MSE bootstrap procedure for non-Poisson outcome variables. This bootstrap scheme for the GMERF operates independently of distributional assumptions and serves as a safeguard against mild deviations from the Poisson distribution.

Although the original MERF approach is tailored for continuous outcome variables, it can theoretically be applied to count data as well. This is achieved by having the MERF iteratively estimate a random forest, assuming the random effects term to be correct, and then estimating the random effects components (and variance components) by decomposing the random forest residuals using a linear mixed model. Therefore, the MERF may provide a valid alternative for modeling count data without relying on explicit distributional assumptions like Poisson or Negative Binomial. We investigate its behavior compared to the GMERF for count data in simulation studies. Additionally, we introduce a modified non-parametric bootstrap method for the MERF, inspired by the approach of \citet{Krennmair2023}, to account for the additional uncertainty due to treating the count variable as continuous.

The paper is organized as follows: Section 2 introduces GMERFs as a methodology that merges random forests with hierarchical modeling to capture dependencies among unit-level (count) observations. Section 2.2 details the process for constructing area-level estimates. To accurately evaluate the MSE of these estimates, Section 3 describes three bootstrap schemes: parametric and non-parametric approaches for GMERFs, and an adjusted non-parametric method for the MERF. Section 4 assesses and compares the effectiveness of the proposed GMERF and MERF methods against the empirical best plug-in predictor (EBPP, \citet{Jiang2003}) under the Poisson generalized linear mixed model through model-based simulation studies. Section 5 applies these methods to estimate area-level means of educational attainment among women in the Mexican state of Guerrero, also addressing the uncertainty of these estimates and discussing the robustness of the modeling techniques. The study concludes in Section 6 with recommendations for further research in the field of SAE. In addition, we provide a design-based simulation study in the Appendix, based on data from Guerrero, to assess the proposed estimators in a close-to-reality environment.

\section{Methodology}\label{sec:2}
This section introduces the GMERF, a semi-parametric unit-level model designed for handling Quasi-Poisson-distributed count outcomes. The model employs a regression forest approach to estimate area-level means, utilizing both unit-level survey and unit-level administrative data. This technique effectively addresses the challenges of modeling count data by combining the flexibility of machine learning methods, which can account for non-linear relationships, with parametric Poisson assumptions.

\subsection{Generalized semi-parametric unit-level model}\label{sec:2.1}
We examine a finite population $P$ divided into $D$ disjunct areas $P_i$ each with a sub-population size of $N_i$, where $i = 1,...,D$ specifies the areas and $N = \sum_{i=1}^{D} N_i$ defines the population size of all areas. The sample $s$ consists of area-specific sub-samples $s_i$ with a total size $n = \sum_{i=1}^{D} n_i $. In contrast, non-sampled observations are denoted as $r_i$ with a size of $N_i-n_i$. We denote individual units within each area as $j \in s_i$ for sampled and $j \in r_i$ for non-sampled observations. The discrete unit-level outcome variable is represented by $y_{ij}$, with information available for $n$ observations within the sample $s$. The vector $\textbf{x}_{ij} = [x_1,x_2,...,x_p]^\intercal$ encompasses $p$ auxiliary covariates, and these auxiliary variables are known for all $N$ units in our population $P$. We assume that $y_{ij}$ follows a generalized semi-parametric unit-level model:
\begin{align}\label{mod1}
	\eta_{ij} & = f(\textbf{x}_{ij})+ \nu_i, \\
	\nu_i &\sim N(0,\sigma^2_{\nu}), \nonumber \\
	E(y_{ij}|\nu_i) & = \mu_{ij} = g(\eta_{ij}), \nonumber
\end{align}
where the fixed part, $f()$, is defined by a random forest and models the conditional mean of the linear predictor ($\eta_{ij}$) given the covariates. The term $\nu_i$ denotes a domain-specific random intercept with its variance $\sigma^2_{\nu}$, capturing small-area variations in the conditional distribution of the linear predictor given $\textbf{x}_{ij}$. $g()$ is a monotonic, differentiable link function that specifies the function of the mean ($\mu_{ij}$) equated to the systematic component. In this work, we assume that the individual $y_{ij}$ values in domain $i$ are independent Poisson random variables with $Var(y_{ij}|\nu_i)=E(y_{ij}|\nu_i)$. Consequently, the mean of the response variable  is connected to the linear predictor via the logarithmic link function: $\mu_{ij} = exp(\eta_{ij})$. If $y_{ij}$ were a continuous variable following a Gaussian distribution, the link function would simplify to the identity function. Thus, in model (\ref{mod1}), the GMERF reduces to the MERF, as proposed by \citet{Krennmair2023}. Furthermore, model (\ref{mod1}) can be seen as a generalized linear mixed model (GLMM) by setting $f(\textbf{x}_{ij})=\textbf{x}^\intercal_{ij}\beta$, where $\beta = [\beta_1,\dots,\beta_p]^\intercal$ represents the regression parameters.

Given that inference using GLMMs presents computational hurdles due to the presence of high-dimensional integrals in the likelihood, which are not amenable to analytical evaluation, we employ an estimation strategy for model (\ref{mod1}) akin to \citet{Frink2024}. This approach utilizes the penalized quasi-likelihood (PQL) method \citep{Breslow1993, Stroup2012} in conjunction with the EM-algorithm \citep{Moon1996} to address the computational challenges inherent in GLMM estimation. More concretely, a weighted MERF pseudo-model is devised, leveraging a linearized target variable $y_{L,ij}=g(\mu_{ij})+(y_{ij}-\mu_{ij})g'(\mu_{ij})$ and weights $w_{ij}=(v_{ij}g'(\mu_{ij})^2)^{-1}$, with $v_{ij}$ being a known variance function. The algorithm follows a doubly iterative process, with micro iterations nested within macro iterations. During each macro-iteration, the linearized response variable and weights are updated. These revised values then serve as the response variable and weights for the subsequent micro-iterations. 
To fit model (1) on survey data, the GMERF algorithm uses initial estimates for $\mu_{ij}$, weights $w_{ij}$ and $y_{L,ij}$ from a generalized linear model (GLM) during the first macro-iteration. In the subsequent micro-iterations, the algorithm (i) estimates the forest function, assuming the random effects to be correct, and (ii) estimates the random effects part using the weighted pseudo-model, assuming the Out-of-Bag (OOB) predictions from the regression forest to be valid. Convergence is monitored by the relative change in the log-likelihood of model (1) between two micro-iterations. It is considered achieved when its relative change is less than a specified number. Once convergence is achieved, the initial values for the next macro-iteration are updated. For the specific case of $y_{ij}$ being Poisson distributed, the initial values are updated as follows:
	\begin{align}\label{mod2}
		\eta_{ij}&= \hat{f}(\textbf{x}_{ij})^{OOB} + \hat{\nu}_i,\\ \nonumber
		\mu_{ij} &= exp(\eta_{ij}), \\ \nonumber
			\end{align}
				\begin{align}\label{mod2}
		y_{L,ij} &= log(\mu_{ij})  \nonumber +\frac{(y_{ij}-\mu_{ij})}{\mu_{ij}},\\ \nonumber
		w_{ij} &= \frac{1}{v_{ij}(\mu_{ij})}, \nonumber
	\end{align}
where $\hat{f}(\textbf{x}_{ij})^{OOB}$ and $\hat{\nu}_i$ reflect their estimated values from the micro-level convergence achieved in the preceding macro-iteration. The predictive performance of our method is significantly influenced by two key tuning parameters: the number of split candidates at each node, which controls the degree of decorrelation, and the number of trees, which together enhance the model’s robustness and accuracy. For further methodological details, we refer to \citet{Frink2024}.

The Poisson distribution is very common for modeling the mean behavior of count outcomes. However, it may underestimate variability in cases of overdispersion, where the observed variance of the response variable exceeds what the Poisson distribution predicts. Overdispersion is a common occurrence in count data analysis and can significantly affect result interpretation. Failure to account for overdispersion in the model can result in overly narrow standard errors and confidence intervals, as well as excessively lenient significance tests, leading to the detection of effects that do not truly exist \citep{VerHoef2007}. One widely used approach to address overdispersion is to employ a Quasi-Poisson model \citep{Gourieroux1984}. In Quasi-Poisson models, $E(y^{qp}_{ij}|\nu_i) = E(y_{ij}|\nu_i) = \mu_{ij} $ and $Var(y^{qp}_{ij}|\nu_i) = \theta \mu_{ij}$, where $\theta$ is a dispersion parameter. The algorithm mentioned above can be directly expanded to incorporate Quasi-Poisson models to address overdispersion concerns. Nevertheless, given our primary focus on predicting means within the machine learning framework and our lesser concern with standard error inference, the Quasi-Poisson model does not yield any discernible advantage within the generalized mixed effect regression forest model. Results for the Quasi-Poisson distribution in terms of GMERFs are available and can be requested from the authors.

\subsection{Domain-level estimator for counts}\label{sec:2.2}
The proposed estimator for the area-level mean is given by:
\begin{align}
	\hat{\eta}_i &= \frac{1}{N_i} \sum_{j=1}^{N_i}\hat{f}(\textbf{x}_{ij}) + \hat{\nu}_i, \nonumber\\
	\hat{\mu}_i &= exp(\hat{\eta}_i)\text{ for $i = 1,\dots, D$.}
\end{align}
For non-sampled areas, the proposed estimator for the area-level mean simplifies to the fixed component obtained from the random forest:
\begin{align*}\label{mod5}
	\hat{\eta}_i &= \frac{1}{N_i} \sum_{j=1}^{N_i}\hat{f}(\textbf{x}_{ij}), \\
	\hat{\mu}_i &= exp(\hat{\eta}_i)\text{.}
\end{align*}
Our proposed estimator $(3)$ offers an advancement over existing SAE methods for count data by addressing issues commonly associated with model selection. This improvement arises from employing the random forest technique, which inherently optimizes model selection to capture higher-order effects and non-linear interactions. Furthermore, this estimator is well-suited for analyzing high-dimensional covariate data, effectively managing scenarios where the number of covariates exceeds the sample size \citep{Hastie_etal2009}. 

However, the GMERF estimator relies on the assumptions of the Poisson distribution due to the linearized target variable and the weights used in the algorithm. In contrast, the MERF, originally designed for continuous outcomes, offers a viable alternative for modeling count data. This approach is suitable for count target variables because it uses continuous residuals as target variables within its inherent linear mixed model. Moreover, the MERF has the advantage of not depending on the assumptions of the Poisson or Negative Binomial distributions, making it a robust option when dealing with severe overdispersion.

\section{Uncertainty estimation}\label{sec:3}
Estimating the MSE of small area estimates poses a substantial challenge \citep{Molina2015}. In this section, we propose two bootstrap schemes to estimate the MSE of the small area estimator introduced in equation (3). The primary distinction between the two bootstrap procedures lies in the mechanism used for generating the bootstrap population. Specifically, the first bootstrap scheme parametrically generates bootstrap realizations of the random effects and explicitly employs the Poisson distribution to create $y_{ij}$. This approach is particularly effective in scenarios where the target variable conforms to a Poisson distribution, as it directly utilizes this distribution. However, count data frequently exhibit overdispersion, which invalidates the assumptions underpinning the Poisson distribution. Consequently, the parametric bootstrap may become unsuitable in such cases. Therefore, we propose adopting a non-parametric bootstrap approach, at least as a supplementary MSE estimation method,  influenced by the work of \citet{Cantoni2019} and \citet{Krennmair2023} for the GMERF. As the approach is independent of the Poisson distribution's assumptions, we expect that the non-parametric bootstrap serves as a safeguard against deviations from the Poisson distribution, especially in the presence of overdispersion. 

Furthermore, we introduce an adjusted non-parametric MSE bootstrap scheme for the MERF with count outcomes. Building on the work of \citet{Chambers2013} and \citet{Krennmair2022}, this bootstrap procedure is adjusted to account for the uncertainty of the estimation, enabling the generation of a discrete count variable in the bootstrap population. Detailed information on this bootstrap method can be found in Appendix A.1.

\subsection{Parametric bootstrap}\label{subsec:3.1}
The procedural steps outlined for the proposed parametric bootstrap of the GMERF closely resemble those detailed in \citet{Frink2024} and can be summarized as follows:
\begin{enumerate}
	\item For $b = 1,\dots, B$:
	\begin{enumerate}
		\item Create bootstrap random effects for each of the $D$ areas by $\nu_i^{(b)} \sim N(0, \hat{\sigma}_{\nu}^2)$. 
		\item Create a bootstrap population of size $N$, by generating values $y^{(b)}_{ij}$ from a Poisson distribution with
		\begin{align*}
			\mu^{(b)}_{ij} = exp \biggl(\hat{f}(\textbf{x}_{ij}) + \nu_i^{(b)} \biggr).
		\end{align*}
		\item Determine the true bootstrap population area means $\mu^{(b)}_{i}$ as  $\frac{1}{N_i}\sum_{j=1}^{N_i} y^{(b)}_{ij}$ for all $i=1,\dots,D$.
		\item For each bootstrap population, select a bootstrap sample that matches the original sample size $n_i$ and estimate  $\hat{f}^{(b)}()$ and $\hat{\nu}_i^{(b)}$. Calculate area-level means $\hat{\mu}^{(b)}_i$.
	\end{enumerate}
	\item Utilizing the $B$ bootstrap samples, the MSE estimator is derived as follows:
	$$\widehat{MSE}_i = \frac{1}{B} \sum_{b=1}^{B} \biggl(\mu_i^{(b)}-\hat{{\mu}}^{(b)}_{i}\biggr)^2.$$
\end{enumerate}

\subsection{Non-parametric bootstrap}\label{subsec:3.2}
The parametric bootstrap may be unsuitable in the presence of overdispersion. To overcome this problem, we propose a non-parametric bootstrap that does not rely on the assumptions of the Poisson distribution. This bootstrap scheme for the GMERF is not only independent of distributional assumptions but also guarantees that a discrete count target variable is generated in the bootstrap population. The proposed bootstrap scheme is described below:

\begin{enumerate}
	\item For given $\hat{f}()$, calculate the marginal pearson residuals $z_{ij} = \frac{y_{ij} - exp(\hat{f}(\textbf{x}_{ij}))}{ \sqrt{exp(\hat{f}(\textbf{x}_{ij})}}$.
	\item Employing the marginal pearson residuals $z_{ij}$, calculate level-2 residuals for each area by $$\bar{z}_{i} = \frac{1}{n_i} \sum_{j=1}^{n_i} {z_{ij}}\enspace\enspace\text{for}\enspace\enspace i=1,...,D.$$
	\item Compute the vector of level-1 residuals by $\hat{z}_{ij} = z_{ij} - \bar{z}_i$. 
Similar to \citet{Krennmair2023}, we adjust the residuals $\hat{z}_{ij}$ to the bias-corrected variance and center them, indicated by $\hat{z}^{c}_{ij}$. Our approach differs in that we use the marginal Pearson residuals of the GMERF, as described in step 1 above, instead of the marginal raw residuals of the MERF for the adjustment process. Level-2 residuals $\bar{z}_{i}$ are additionally adjusted for the estimated variance $\hat{\sigma}^2_{\nu}$ and centered, indicated by $\bar{z}^c$.
	\item For $b=1,...,B$:
	\begin{enumerate}
		\item Independently draw samples with replacement from the scaled and centered level-1 and level-2 residuals:
		\begin{eqnarray}
			\nonumber z_{ij}^{(b)}=\text{srswr}(\hat{z}^c_{ij},N)\enspace\enspace \text{and}\enspace\enspace \bar{z}^{(b)}_i=\text{srswr}(\bar{z}^c,D).
		\end{eqnarray}
		\item[(b)] Calculate  $\mu^{(b)}_{ij} = exp(\hat{f}(\textbf{x}_{ij}) + \bar{z}_i^{(b)})$ and the corresponding $\tilde{y}^{(b)}_{ij} = \mu^{(b)}_{ij} + \sqrt{\mu^{(b)}_{ij}} \times z_{ij}^{(b)}$ for $j=1,\dots,N$. Note that $\tilde{y}^{(b)}_{ij}$ is defined on a continuous scale.
		\item[(c)] To obtain a count target variable, we match $\tilde{y}^{(b)}_{ij}$ with the set of estimated unit-level predictors from the sample $\hat{\mu}_{t} = exp(\hat{f}(\textbf{x}_{ij}) + \hat{\nu}_i)$ by finding the corresponding index $\tilde{t}$ solving $$\min_{t\in s} |\tilde{y}^{(b)}_{ij} - \hat{\mu}_t|.$$
		\item[(d)] Define the bootstrap population by $y^{(b)}_{ij} = y_{\tilde{t}}$ for $j=1,\dots,N$ and calculate the true bootstrap population mean $\mu_i^{(b)}$ for $i = 1,...,D$.
		\item[(e)]For each bootstrap population, select a bootstrap sample that matches the original sample size $n_i$ and estimate $\hat{f}^{(b)}()$ and $\hat{\nu}_i^{(b)}$. Obtain estimates for the mean $\hat{\mu}_i^{(b)}$.
	\end{enumerate}
	\item Using the $B$ bootstrap samples, the MSE estimator is computed as follows:
	$$\widehat{MSE}_i = \frac{1}{B} \sum_{b=1}^{B} \biggl(\mu_i^{(b)}-\hat{{\mu}}^{(b)}_{i}\biggr)^2.$$
\end{enumerate}
Both bootstrap schemes we described in this section are empirically evaluated in Section 4.

\section{Model-based simulation study}\label{sec:4}
The use of model-based simulations enables a controlled empirical assessment of our proposed methods for point and uncertainty estimates. We compare the point and uncertainty estimates for domain-level means derived from the GMERF model (\ref{mod1}) with those obtained from two competing models. In particular, we examine the performance of GMERFs in comparison to the EBPP, which is based on a Poisson GLMM: $$\hat{\mu}^{EBPP}_i=\frac{1}{N_i}\Biggl(\sum_{j \in s_i} y_{ij}+\sum_{j \in r_i} \hat{\mu}_{ij}\Biggr),$$
where $\hat{\mu}_{ij} = exp(\textbf{x}_{ij}^\intercal \hat{\beta}+\hat{\nu}_i)$ \citep{Jiang2003}.  By contrasting the performance of this linear-based competitor (on the linear predictor sale) with our more flexible approach, which incorporates semi-parametric and non-linear modeling, we aim to showcase the advantages of the GMERF methodology. Additionally, we compare the GMERF method with the MERF (for originally continuous outcomes): $$\hat{\mu}^{MERF}_{i}=\frac{1}{N_i} \sum_{j=1}^{N_i}\hat{f}(\textbf{x}_{ij}) + \hat{\nu}_i.$$ The aim of comparing both machine learning methods is to highlight the advantages of GMERF when the Poisson distribution is applicable or only moderately violated, and to demonstrate the suitability of MERF for count data in situations of severe overdispersion.

The simulation setup involves a finite population $U$ with a total size of $N=50,000$, divided into $D=50$ disjoint areas $U_1, ..., U_D$ each having $N_i=1,000$ units. We generate samples using stratified random sampling, where the $50$ small areas are treated as strata. This approach yields a total sample size of $n = \sum_{i=1}^D n_i = 921$. The number of sampled units per area varies from $8$ to $29$, with a median of $18$. These sample sizes are consistent with the area-level sample sizes found in the application discussed in Section 5.

We consider four scenarios denoted as \textit{Normal-Poisson}, \textit{Interaction-Poisson}, \textit{NB3}, \textit{NB1} and repeat each scenario independently $M=500$ times. The comparison of competing model-estimates under these four scenarios allows us to examine the performance under two major dimensions: Firstly, the presence of overdispersed data delineated by the Negative Binomial distribution and secondly, the presence of unknown non-linear interactions between covariates on the linear predictor scale. 

We initially establish a baseline scenario, \textit{Normal-Poisson}, for Poisson GLMMs generating Poisson-distributed outcomes without interactions or quadratic terms. 
Since the assumption of linearity in the model is satisfied, our goal is to demonstrate that GMERFs perform similar to their linear counterparts in the reference scenario. In contrast, the \textit{Interaction-Poisson} scenario introduces a more intricate model, incorporating quadratic terms and interactions at the linear predictor level. This scenario aims to underscore the benefits of non-linear modeling approaches. Overdispersion frequently complicates the analysis of count data. To more realistically mirror these situations, we apply a Negative Binomial distribution (with scale parameter $3$ and $1$) in scenarios \textit{NB3} and \textit{NB1}. These scenarios utilize complex models with interaction effects but vary according to the scale parameter $s$, which influences the degree of overdispersion. A smaller $s$ value indicates increased overdispersion. Further information on the data generation process for each scenario is detailed in Table \ref{tab:MB1}.

We evaluate the point estimates for the area-level means using two quality measures: bias and root mean squared error (RMSE). To assess the proposed MSE estimators, we examine the relative bias of the root mean squared error (RB-RMSE) and the relative root mean squared error of the RMSE:
\begin{align}
	\nonumber	BIAS_i &= \frac{1}{M} \sum_{m=1}^{M} \left(\hat{\mu}^{(m)}_i - \mu^{(m)}_i\right)\\\nonumber
	\textcolor{black}{RMSE_i} &= \textcolor{black}{\sqrt{\frac{1}{M} \sum_{m=1}^{M} \left(\hat{\mu}^{(m)}_i - \mu^{(m)}_i\right)^2}}\\\nonumber
	RB\text{-}RMSE_i &=\frac{\sqrt{\frac{1}{M} \sum_{m=1}^{M} MSE^{(m)}_{est_i}} - RMSE_{i}}{RMSE_{i}}\\\nonumber
	RRMSE\text{-}RMSE_i &= \frac{\sqrt{\frac{1}{M} \sum_{m=1}^{M} \left(\sqrt{MSE^{(m)}_{est_i}} - RMSE_{i}\right)^2}}{RMSE_{i}},
\end{align}
where $\hat{\mu}^{(m)}_i$ denotes the estimated mean for area $i$ derived from any of the aforementioned methods, and $\mu^{(m)}_i$ represents the true mean for area $i$ in simulation round $m$. The estimation of $MSE_{est_i}^{(m)}$ is carried out using the proposed bootstrap methods detailed in Section \ref{sec:3} and Appendix A.1.

\begin{table}
	\centering
	\captionsetup{justification=centering,margin=1.5cm}
	\caption{Model-based simulation scenarios.}
	\resizebox{\textwidth}{!}{\begin{tabular}{rlccccccc}
			\toprule
			{Scenario} & {Linear predictor} & {Mean} & {$y$} & {$x_1$} & {$x_2$} & {$\nu$} \\ \midrule
			\textit{Normal-Poisson}  & $ \eta = 2+x_1+x_2+\nu$& $\mu=exp(\eta)$& $ Pois(\mu)$& $U(-1,1)$ & $N(-1,1)$  & $N(0,0.3^2)$\\
			\textit{Interaction-Poisson}  & $ \eta = 2+2x_1x_2+x_2^2+\nu $ &$\mu=exp(\eta)$& $ Pois(\mu)$& $U(-1,1)$ & $N(-1,1)$  & $N(0,0.3^2)$\\
			\textit{NB3}  & $ \eta = 2+2x_1x_2+x_2^2+\nu $&$\mu=exp(\eta)$& $ NB(\mu,3)$ &$U(-1,1)$ & $N(-1,1)$  & $N(0,0.3^2)$\\
			\textit{NB1}  & $ \eta = 2+2x_1x_2+x_2^2+\nu $&$\mu=exp(\eta)$& $ NB(\mu,1)$ & $U(-1,1)$ & $N(-1,1)$  & $N(0,0.3^2)$  \\\bottomrule
	\end{tabular}}
	\label{tab:MB1}
\end{table}

To conduct the model-based simulation, we utilize \emph{R} \citep{R_language}. The EBPP estimates are generated using the \emph{lme4} package \citep{Bates_etal2015}. For the proposed GMERF and MERF methods, we use the \emph{ranger} package \citep{Wright17} alongside \emph{lme4} \citep{Bates_etal2015}. To ensure the algorithms converge, we apply a precision of $1e^{-5}$ for the relative change in the log-likelihood criterion (applicable to the MERF algorithm as well) and a precision of 0.001 for the relative change in $\hat{\eta}$. 

Table 2 reports the empirical bias and the RMSE of each method across the four scenarios. In the \textit{Normal-Poisson} scenario the EBPP, MERF, and GMERF estimators show varying degrees of positive bias, with the median bias magnitude increasing in that order. For the \textit{Interaction - Poisson} data-generating process, biases not only increase but also change sign for all methods, indicating more complex error dynamics. In the Negative Binomial scenarios, designed to examine model performance under conditions of overdispersion, the MERF demonstrates a lower bias in comparison to the GMERF. The GMERF shows the highest bias across these scenarios, likely due to its reliance on the approximation method in its algorithm.

\begin{table}[!h]
	\footnotesize
	\centering
	\captionsetup{justification=centering,margin=1.5cm}
	\caption{Mean and median of bias and RMSE over areas for point estimates.}
	\begin{tabular}{lcccccccc}
		\\[-1.8ex]\hline
		\hline \\[-1.8ex]
		&\multicolumn{2}{c}{\textit{Normal-Poisson}} &\multicolumn{2}{c}{\textit{Interaction-Poisson}}&\multicolumn{2}{c}{\textit{NB3}}&\multicolumn{2}{c}{\textit{NB1}} \\
		\hline \\[-1.8ex]
		& Median & Mean & Median & Mean & Median & Mean & Median & Mean\\
		\hline \\[-1.8ex]
		\multicolumn{8}{l}{Bias}\\
		\hline \\[-1.8ex]
		EBPP &  $0.0054$ & $0.0123$& $-0.0895$ & $-0.0887$ &  $-0.0609$&  $-0.0489$&  $-0.0894$& $-0.1013$\\
		GMERF & $0.0774 $ & $0.0834$& $-0.1376$ & $-0.1587$ & $0.4432$ & $0.4981$ & $3.1600$ & $3.1510$ \\
		MERF &$0.0075$ & $0.0099$ & $-0.0628$ & $-0.0628$ & $0.0830$ & $0.0592$ & $0.0675$ & $0.0750$ \\
		\hline \\[-1.8ex]
		\multicolumn{8}{l}{RMSE}\\
		\hline \\[-1.8ex]			
		EBPP & $1.0910$ & $1.1525$ & $1.5450$ & $1.6360$ &$4.3740$& $4.5650$ &  $7.3670$& $7.5960$\\
		GMERF & $1.2449$ & $1.2979$ & $1.4300$ & $1.4860$ & $4.2140$ & $4.3490$ & $8.2460$ & $8.8140$\\
		MERF& $1.3610$ & $1.4370$ & $1.9300$ & $2.0120$ & $4.1360$ & $4.2170$ & $5.6160$ & $5.6930$ \\\hline \\[-1.8ex]
	\end{tabular}
	\label{tab:mod_biasrmse_tab}
\end{table}

Figure 1, along with Table 2, evaluates the RMSE for each estimation method across various scenarios. In the \textit{Normal-Poisson} scenario, EBPP outperforms both MERF and GMERF. This suggests that EBPP's adherence to the model's fixed effects leads to more accurate estimates when the assumptions of the Poisson distribution are met. In the more complex \textit{Interaction-Poisson} scenario, MERF shows a higher RMSE compared to EBPP, indicating lower accuracy. MERF's performance suggests a greater sensitivity to deviations of distribution assumptions compared to violations of linearity assumptions, as EBPP remains more efficient under the Poisson assumption with non-linear relationships in the fixed effects part of the model. Notably, GMERF's point estimates surpass those of EBPP, highlighting its effectiveness in complex models that involve interactions and non-linear relationships. In scenarios incorporating a Negative Binomial distribution, MERF's performance tends to improve as the degree of overdispersion increases. This superiority (especially in the \textit{NB1} scenario) highlights MERF's ability to effectively handle overdispersed count data, benefiting from its flexibility in modeling both continuous and discrete targets without strict reliance on Poisson model assumptions. Overall, the insights from Table 2 and Figure 1 indicate that GMERF delivers competitive performance when the Poisson distribution assumptions are moderately met, whereas MERF excels in scenarios where these assumptions are severely violated.

\begin{figure}[!htb]
	\centering
	\captionsetup{justification=centering,margin=1.5cm}
	\includegraphics[width=1\linewidth]{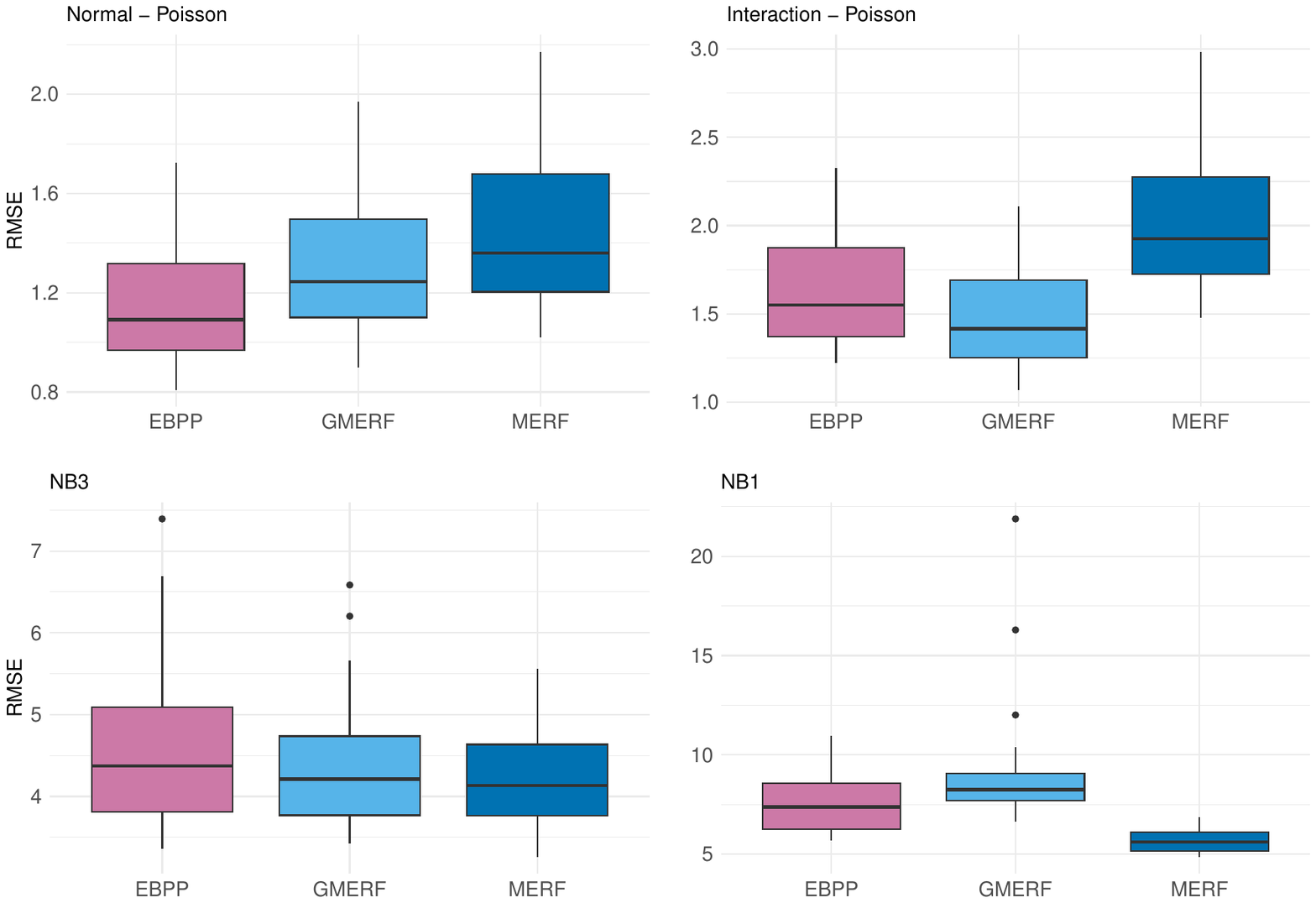}
	\caption{Empirical RMSE comparison of point estimates for area-level averages under four scenarios.}
	\label{fig:mod_point_fig}
\end{figure}

We now assess the performance of the MSE estimators presented in Section 3 and Appendix A.1 using $B=200$ bootstrap replications. Table $3$ displays the performance of the three bootstrap procedures, evaluated using RB-RMSE and RMSE-RMSE. Specifically, the proposed MSE estimators show reasonably low to moderate relative bias in terms of both mean and median values across all scenarios. The parametric bootstrap estimator (GMERF P) performs similarly in simpler scenarios, such as  \textit{Normal-Poisson}, and in more complex scenarios like \textit{Interaction-Poisson}. Notably, we do not provide results for Negative Binomial settings with this estimator due to its reliance on Poisson distribution assumptions, which can lead to exaggerated MSE estimates when these assumptions are violated. These results are available upon request from the authors. The non-parametric bootstrap estimator for GMERF (GMERF NP) exhibits slight variability in relative bias across different scenarios. The adjusted non-parametric bootstrap estimator for MERF (MERF NPC) consistently shows low bias and maintains precision across various scenarios. Notably, GMERF NP performs comparably to MERF NPC in scenarios involving a Negative Binomial distribution in terms of RB-RMSE. While Table 3 does not directly convey the area tracking properties of the estimated RMSE versus the empirical RMSE, Figure 2 provides further insights. Based on the tracking properties, we conclude that using the parametric and non-parametric bootstraps for estimating the MSE appear to have appealing properties regarding bias and stability.

\begin{table}[!h]
	\footnotesize
	\centering
	\captionsetup{justification=centering,margin=1.5cm}
	\caption{Performance of bootstrap MSE estimators in model-based simulation: mean and median of RB-RMSE and RRMSE-RMSE over areas.}
	\begin{tabular}{lcccccccc}
		\\[-1.8ex]\hline
		\hline \\[-1.8ex]
		 &\multicolumn{2}{c}{\textit{Normal-Poisson}} &\multicolumn{2}{c}{\textit{Interaction-Poisson}}&\multicolumn{2}{c}{\textit{NB3}}&\multicolumn{2}{c}{\textit{NB1}} \\
		\hline \\[-1.8ex]
		 & Median & Mean & Median & Mean &  Median & Mean& Median & Mean \\
		\hline \\[-1.8ex]
		\multicolumn{8}{l}{RB-RMSE[\%]}\\
		\hline \\[-1.8ex]
		GMERF P &  $3.57$ & $3.45$& $-0.36$ & $5.33$ &  &  &  & \\
		GMERF NP & $8.23 $ & $8.21$& $-2.24$ & $-1.29$  & $5.81$ & $6.71$ &$-8.54$ &$2.21$\\
		MERF NPC &$1.97$ & $2.03$ & $0.28$ & $0.48$  & $3.95$ & $4.21$ &$-0.83$ & $-1.21$\\
		\hline \\[-1.8ex]
		\multicolumn{8}{l}{RRMSE-RMSE[\%]}\\
		\hline \\[-1.8ex]			
		GMERF P & $10.16$ & $10.56$ & $47.91$ & $57.39$ &&  &  &\\
		GMERF NP & $23.01 $ & $26.26$ & $24.42$ & $24.77$ & $42.79$ & $47.21$ &$60.75$ & $74.41$\\
		MERF NPC & $7.90$ & $8.36$ & $12.86$ & $13.29$ & $15.27$ & $15.67$ & $23.97$ & $24.35$\\\hline \\[-1.8ex]
	\end{tabular}
	\label{tab:mod_mse_tab}
\end{table}

\begin{figure}[!htb]
	\centering
	\captionsetup{justification=centering,margin=1.5cm}
	\includegraphics[width=1\linewidth]{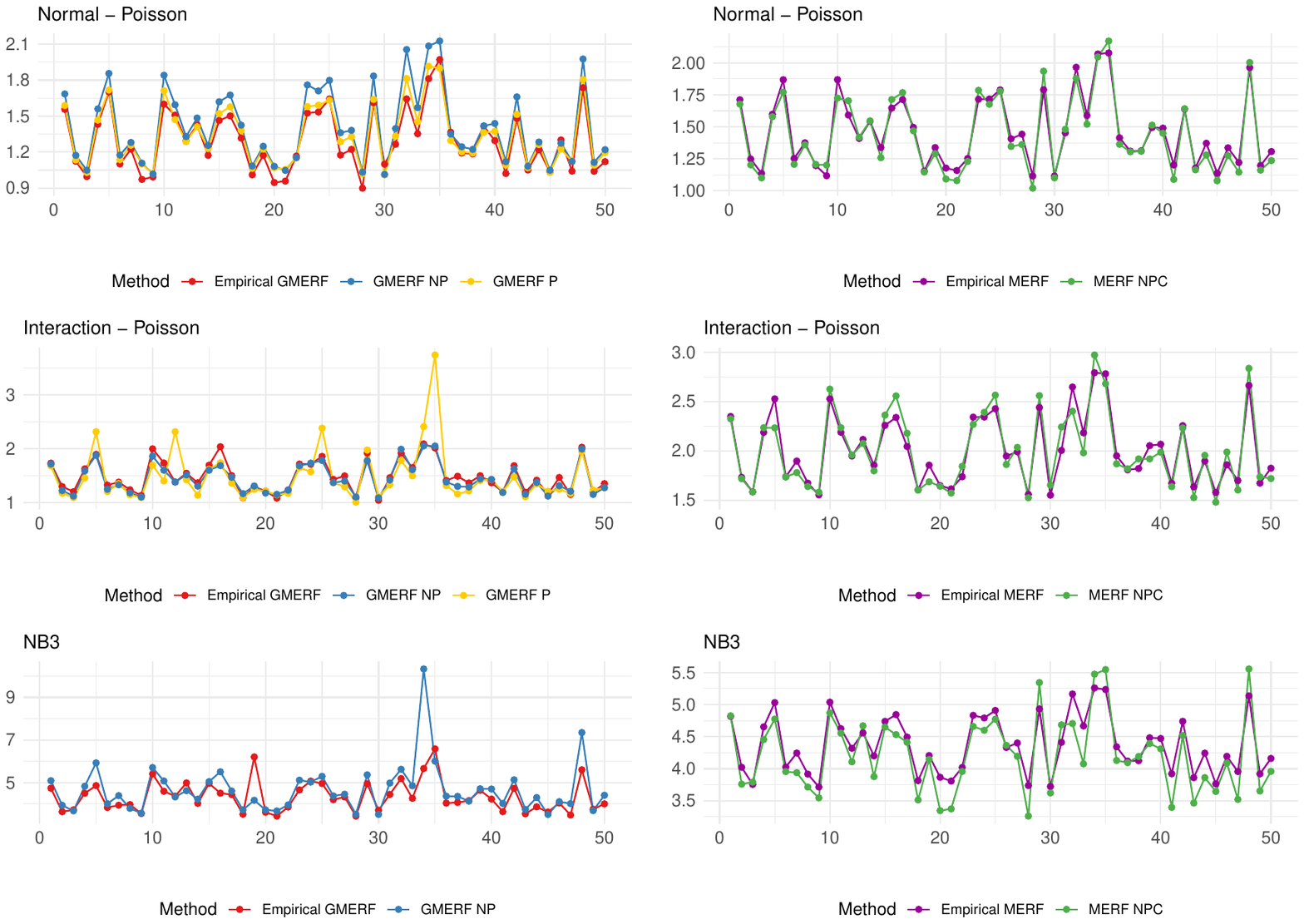}
	\caption{Empirical and bootstrapped area-level RMSEs for three scenarios.}
	\label{fig:mod_mse_fig}
\end{figure}

\section{Application to education data from Guerrero-Mexico}\label{sec:5}

In this section, we describe the application of our SAE methods to estimate the average number of school years completed by women in the municipalities of Guerrero, Mexico, based on data from $2010$.

We combine two primary data sources: the Mexican household income and expenditure survey ENIGH (Encuesta Nacional de Ingreso y Gastos de los Hogares) and census microdata provided by the National Institute of Statistics and Geography (Instituto Nacional de Estadística y Geografía). The ENIGH survey encompasses data from $1,445$ households spread across $40$ municipalities in Guerrero, while the census dataset includes information on $112,265$ households from all $81$ municipalities in the state. The survey samples are notably varied, with individual municipality samples ranging from a minimum of $8$ households to a maximum of $458$, and a median of $21$ households per municipality. This disparity in sample sizes results in $41$ municipalities being left out of the survey sample, highlighting a significant challenge in applying SAE methods effectively. Table \ref{tab:Apdetails} summarizes these domain-specific data characteristics, providing an overview of the coverage and distribution of data points across the municipalities. 

\begin{table}[!htb]
	\centering
	\captionsetup{justification=centering,margin=1.5cm}
	\caption{Summary statistics on in- and out-of-sample areas: area-specific sample size of census and survey data.}
	\begin{tabular}{@{\extracolsep{5pt}} lcccccccc}
		\\[-1.8ex]\hline
		\hline \\[-1.8ex]
		Municipalities &\multicolumn{2}{c}{Total}&\multicolumn{2}{c}{In-sample}&\multicolumn{2}{c}{Out-of-sample}\\
		&\multicolumn{2}{c}{81} & \multicolumn{2}{c}{40} & \multicolumn{2}{c}{41} \\ \hline
		\hline \\[-1.8ex]
		& Min. & Q1 & Median & Mean & Q3 & Max. \\
		\hline
		Survey area sizes & 8.00 & 15.00 & 21.00 & 36.12 & 32.00 & 458.00 \\
		Census area sizes & 361.00 & 619.00 & 833.00 & 1,386.00 & 1,656.00 & 6,297.00 \\
		\hline
	\end{tabular}
	\label{tab:Apdetails}
\end{table}

The implementation of SAE methodologies involves the development of a working model using survey data, which is subsequently refined with census or administrative data. In the context of Poisson GLMMs, such as the EBPP approach, selecting the appropriate variables is crucial. We apply the Akaike information criterion to identify the optimal model for predicting the average number of school years completed by women. This process resulted in the selection of $21$ out of $39$ potential covariates for our final Poisson GLMM model.

\begin{figure}[!htb]
	\centering
	\captionsetup{justification=centering,margin=1.5cm}
	\includegraphics[width=1\linewidth]{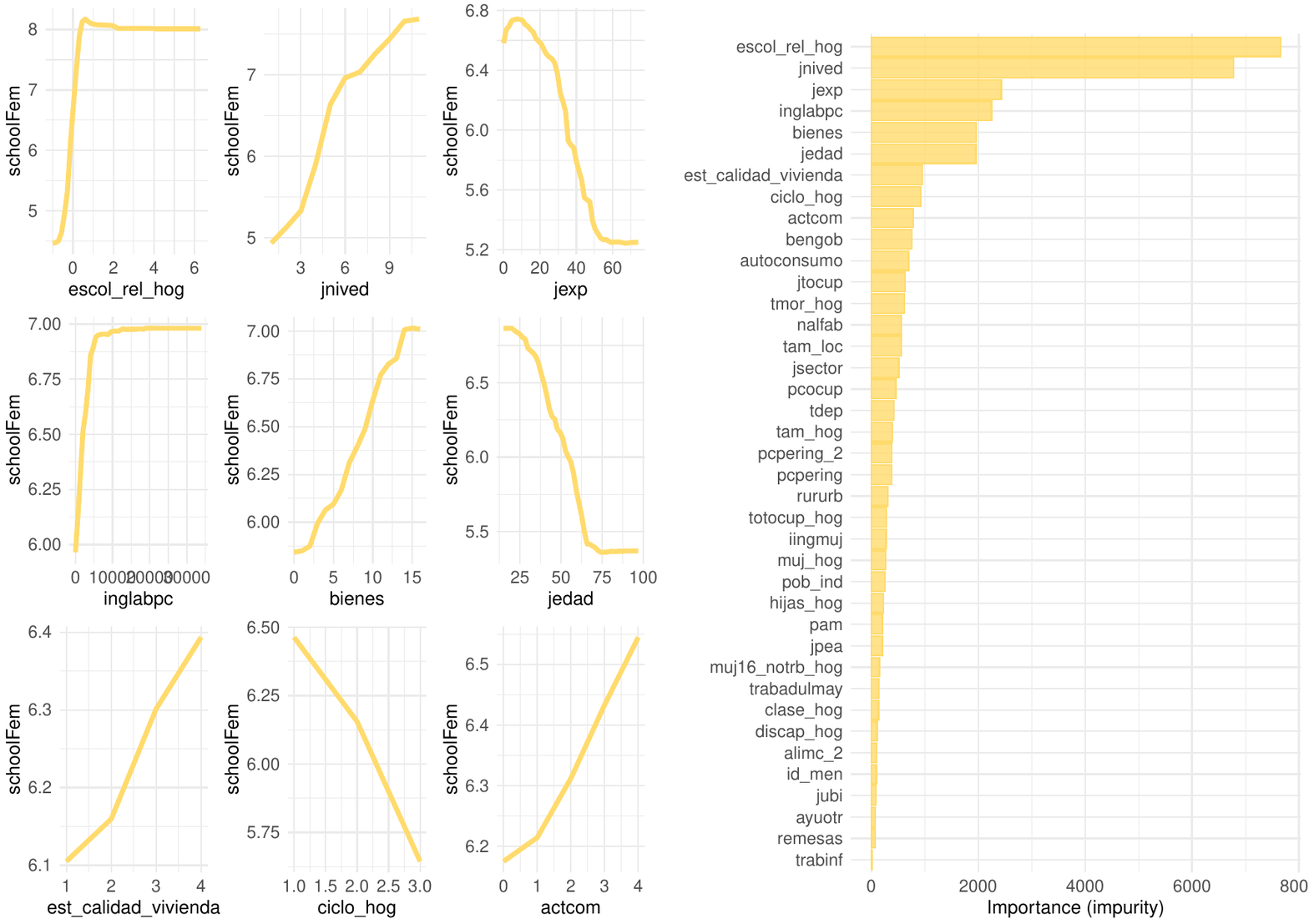}
	\caption{Visual diagnostics on predictive relations  between dependent variable \textit{schoolFem} and predictors. Partial dependency plots (left-hand side); variable importance plot (right-hand side).}
	\label{fig:vippdp}
\end{figure}

In contrast to the explicit variable selection required for Poisson GLMMs, random forests implement an implicit model selection process \citep{Breiman2001}. Figure \ref{fig:vippdp} illustrates the insights gained from this method through partial dependence plots (PDP) and variable importance plots (VIP) \citep{Greenwell2017, Greenwell2020}. The PDP quantifies the marginal effect of specific predictors on the target variable, highlighting the non-linear dynamics between predictors and outcomes. The VIP ranks the significance of each predictor based on the mean decrease in impurity (variance), which is aggregated from the number of splits across all trees that involve the predictor. Notably, key predictors such as years of working experience (jexp), household income (inglabpc) and the level of education in relation to average (escol\_rel\_hog) emerge as critical for modeling the educational attainment of women. Overall, the analysis in Figure \ref{fig:vippdp} underscores the complex interplay between predictors and the target variable, affirming the value of advanced modeling techniques in uncovering these relationships in the context of SAE.

\begin{figure}[!htb]
	\centering
	\captionsetup{justification=centering,margin=1.5cm}
	\includegraphics[width=1\linewidth]{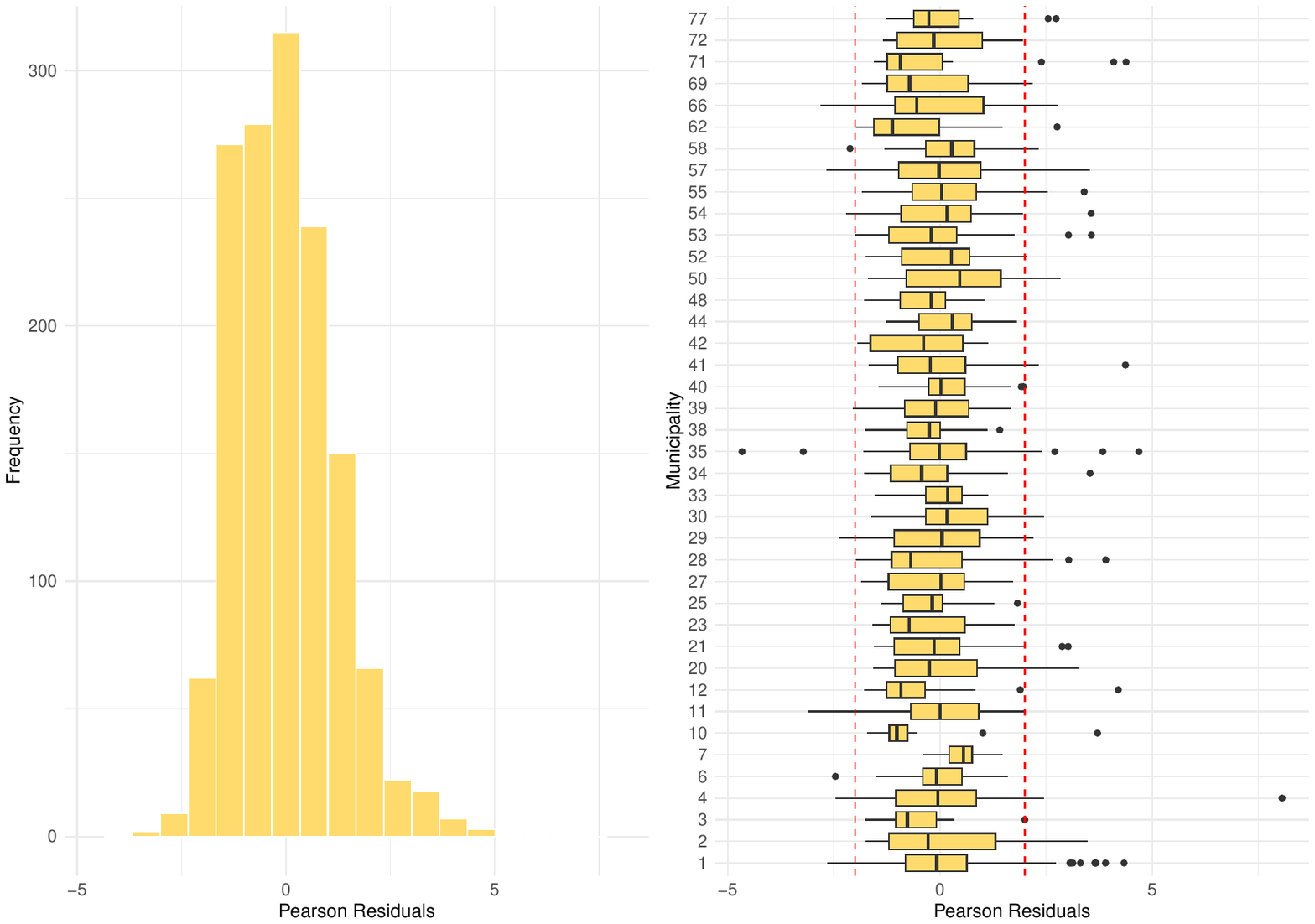}
	\caption{Model fit diagnostics for the Poisson GLMM: Histogram of Pearson residuals (left) and box-plots of Pearson residuals by municipalities (right).}
	\label{fig:pearsonresid}
\end{figure}

Figure \ref{fig:pearsonresid} presents an analysis of Pearson residuals from the Poisson GLMM applied to the estimation of school years for women in Guerrero municipalities. The histogram displays a positive skew in the residuals and the presence of some large values. This pattern is further evidenced by the plot showing the distribution of residuals across municipalities, with several exhibiting numerous positive residuals. The red dashed lines in the right plot of Figure 4 indicate the values -2 and 2, beyond which the Pearson residuals may suggest a poor model fit. Several residuals can be seen exceeding these values. These observations suggest a potential issue with overdispersion - a critical aspect considering the equidispersion assumption of Poisson models. To verify this, we conducted overdispersion tests comparing Poisson and Negative Binomial models. Both the Likelihood Ratio Test (LRT) and Dean's PB test indicate significant overdispersion, with LRT yielding a statistic of $67.471$ and $p<2.2e^{-16}$, and Dean's PB test showing $6.840$ with $p=3.953e^{-12}$.

\begin{figure}[!htb]
	\centering
	\captionsetup{justification=centering,margin=1.5cm}
	\includegraphics[width=1\linewidth]{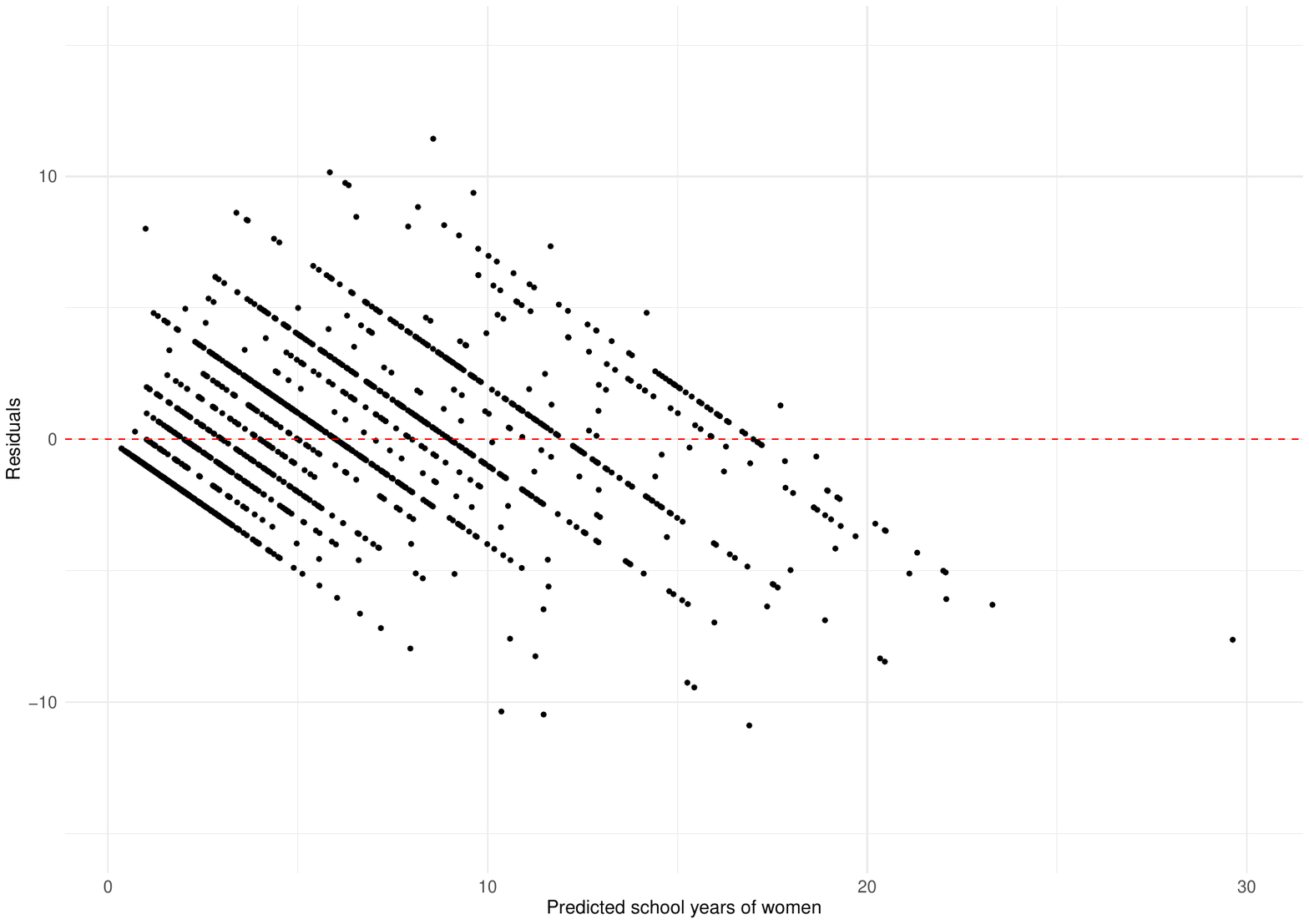}
	\caption{Model fit diagnostics for the Poisson GLMM: Raw residuals vs. predicted values.}
	\label{fig:residpred}
\end{figure}

Additionally, Figure \ref{fig:residpred} illustrates a plot of raw residuals against fitted values, revealing a distinct pattern across the range of fitted values. This pattern of residuals underscores potential misspecification in the Poisson GLMM. Given these findings, we argue that the model's limitations may warrant the exploration of alternative, more flexible modeling approaches. Consequently, we estimate both the GMERF and the MERF. These data-driven semi-parametric models could potentially offer an improved fit and robustness over the traditional parametric EBPP, which is based on a Poisson GLMM.


Figure 6 showcases the results from three different estimation methods applied. These methods include direct estimation, which is feasible for 40 of the 81 municipalities, as well as the model-based approaches provided by GMERF and MERF. The direct estimates serve as a baseline, feasible only for municipalities with sufficient data. Conversely, GMERF and MERF extend our insights into regions where direct data are lacking, thereby enhancing our understanding of regional disparities in educational attainment. All three methods illustrate clear regional differences in educational outcomes among the municipalities. However, the point estimates from GMERF and MERF are similar, suggesting that both model-based methods provide consistent estimates despite their methodological differences. Furthermore, the maps provide useful information on the geographical distribution of the average number of years of schooling for women.

\begin{figure}[!htb]
	\centering
	\captionsetup{justification=centering,margin=1.5cm}
	\includegraphics[width=1\linewidth]{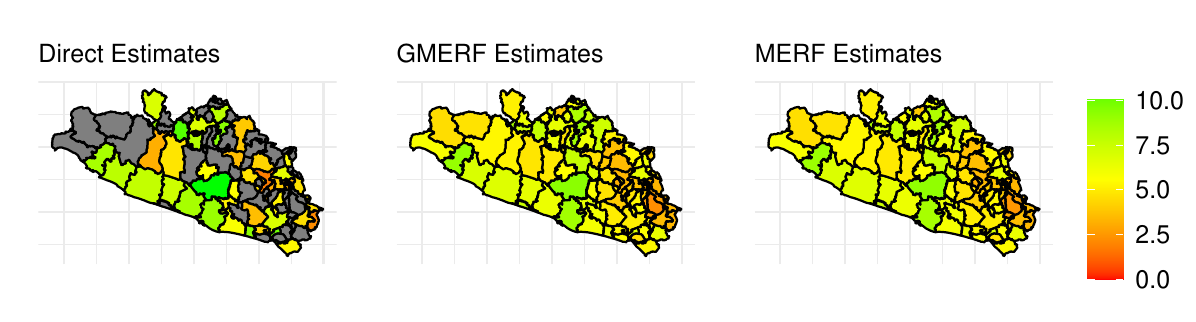}
	\caption{Estimated school years of women for the state of Guerrero based on direct estimates, GMERF and MERF.}
	\label{fig:map}
\end{figure}

Figure 7 highlights the variability in the estimation accuracy for the average number of school years for women, as measured by the coefficients of variation (CV). We use the calibrated bootstrap method from the \textit{R} package \textit{emdi} \citep{Kreutzmann_etal2019} for direct estimates, and the non-parametric bootstrap procedures outlined in Section 3 and Appendix A.1 for GMERF and MERF, each involving $B = 200$ bootstrap replications. Given the established presence of overdispersion, we opted to forgo the use of the parametric bootstrap in this application. For both in-sample and out-of-sample domains, the comparison between GMERF and MERF reveals no notable differences in precision, suggesting comparable performance across these model-based methods.

The observed similarities in CVs across different machine learning models and bootstrap methods could be attributed to the underlying strength of overdispersion in the data, as evidenced by a dispersion ratio of $1.50$ in the estimated Poisson GLMM. This corresponds closely with the scale parameter of $3.071$ in the Negative Binomial distribution used in the analysis, which aligns well with the model-based simulation results for the \textit{NB3} setting (scale parameter of 3) focusing solely on in-sample domains. It seems that there are no substantial differences between GMERF and MERF in a data scenario that shows only slight to moderate overdispersion. 

Further validation of these findings is provided in Appendix A.2, where a design-based simulation study enhances the reliability of the point and MSE estimators discussed here, allowing for a more detailed exploration of their performance.

\begin{figure}[!htb]
	\centering
	\captionsetup{justification=centering,margin=1.5cm}
	\includegraphics[width=1\linewidth]{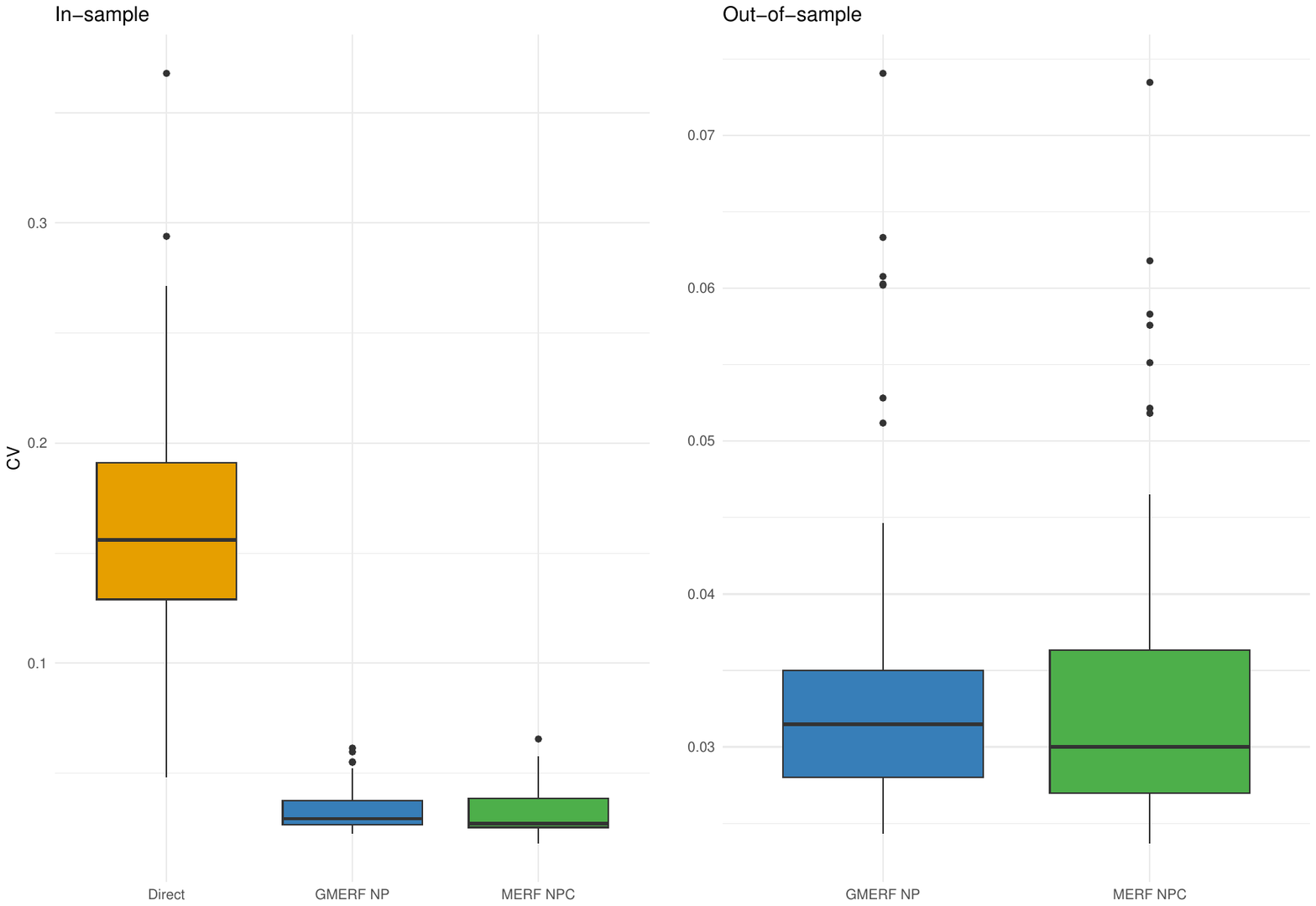}
	\caption{Area-specific CVs for Means for in- and out-of-sample areas.}
	\label{fig:cv}
\end{figure}

\section{Conclusion}\label{sec:6}
In this paper, we introduce generalized tree-based machine learning methods for estimating disaggregated, small area means for count data. Additionally, we explore the impact of overdispersion on the performance of these estimation methods. Our findings reveal that the MERF, which operates independently of distributional assumptions with respect to the count data, performs superior under conditions of severe overdispersion compared to the GMERF. Conversely, the GMERF shows better performance when the assumptions of the Poisson distribution are met or only moderately violated.

In Section 2, we extend the GMERF procedure \citep{Frink2024} to accommodate count data and discuss generalized semi-parametric mixed models at the unit level, treating Poisson GLMM-based SAE methods such as EBPP as special cases. Furthermore, the MERF approach \citep{Krennmair2023}, initially developed for continuous data, is adapted for discrete outcome variables, exploring potential advantages and limitations, especially in relation to overdispersion. We present also three bootstrap methods for evaluating point estimators for area-level means: a parametric and a non-parametric MSE bootstrap procedure for the GMERF, and an adjusted non-parametric MSE bootstrap procedure for the MERF (originally for continuous outcomes). The efficacy of point and MSE estimates is assessed through model-based simulations. Additionally, Section 5 applies these methods to data from the Mexican state of Guerrero.

The simulations in Section 4 demonstrate that the GMERF and MERF point estimates surpass traditional methods when dealing with non-linear interactions between covariates at the linear predictor level. Notably, in scenarios of overdispersion, the MERF remains robust without dependence on Poisson distribution assumptions, outperforming generalized models. The MSE-bootstrap schemes proposed are validated as reliable based on their performance in our simulations. The practical application in Guerrero and further design-based simulations in our Appendix also support the effectiveness of these approaches.

From a methodological standpoint, the scope for further research into the application of GMERF in SAE is broad and promising. One potential direction is to adapt GMERF to model non-linear indicators, such as quantiles. This could involve applying the smearing technique introduced by \citet{Chambers1986} and further developed for machine-learning methods by \citet{KrennmairSchTz2023}. Furthermore, extending the GMERF to accommodate the Negative Binomial distribution by incorporating an additional dispersion parameter into the model could address issues of overdispersion more effectively than models based solely on the Poisson distribution. This modification would potentially allow GMERF to provide more accurate estimates for count data characterized by high variability, making it a more versatile tool in the field of SAE. These expansions not only enhance the flexibility and applicability of GMERF but also open up new possibilities for addressing complex problems in statistical estimation.

\section*{Acknowledgements}
The authors are grateful to CONEVAL for providing the data used in empirical work. The views set out in this paper are those of the authors and do not reflect the official opinion of CONEVAL. The numerical results are not official estimates and are only produced for illustrating the methods. The authors are grateful for the computation time provided by the HPC service of the Freie Universität Berlin.

\clearpage

\bibliographystyle{apacite}		
\bibliography{./GMERF_paper_refs_1}

\clearpage
\begin{appendices}
	\renewcommand{\thesection}{A}

\section{}

\subsection{Non-parametric bootstrap for MERF with count data}
\begin{enumerate}
	\item For given $\hat{f}()$, compute the marginal residuals $z_{ij} = y_{ij} -\hat{f}(\textbf{x}_{ij})$.
	\item Utilizing the marginal residuals $z_{ij}$, determine level-2 residuals for each area by $$\bar{z}_{i} = \frac{1}{n_i} \sum_{j=1}^{n_i} {z_{ij}}\enspace\enspace\text{for}\enspace\enspace i=1,...,D$$
	\item Derive the vector of level-1 residuals using $\hat{z}_{ij} = z_{ij} - \bar{z}_i$. Following \citet{Krennmair2023}, the residuals are adjusted to the bias-corrected variance and centered, indicated by $\hat{z}^{c}_{ij}$. Level-2 residuals $\bar{z}_i$ are additionally adjusted for the estimated variance $\hat{\sigma}_{\nu}^2$ and centered, indicated by $\bar{z}^{c}$.
	\item For $b=1,...,B$:
	\begin{enumerate}
		\item Independently draw samples with replacement from the scaled and centered level-1 and level-2 residuals:
		\begin{eqnarray}
			\nonumber z_{ij}^{(b)}=\text{srswr}(\hat{z}^c_{ij},N)\enspace\enspace \text{and}\enspace\enspace \bar{z}^{(b)}_i=\text{srswr}(\bar{z}^c,D).
		\end{eqnarray}
		\item[(b)] Compute  $\eta^{(b)}_{ij} = \hat{f}(\textbf{x}_{ij}) + \bar{z}_i^{(b)}+ z_{ij}^{(b)}$.
		\item[(c)] To obtain a count target variable, we match $\eta^{(b)}_{ij}$ with the set of estimated unit-level predictors from the sample $\hat{\eta}_{t} = \hat{f}(\textbf{x}_{ij}) + \hat{\nu}_i$ by finding the corresponding index $\tilde{t}$ solving $$\min_{t\in s} |\eta_{ij}^{(b)} - \hat{\eta}_t|.$$
		\item[(d)] Define the bootstrap population by $y^{(b)}_{ij} = y_{\tilde{t}}$ for $j=1,\dots,N$ and calculate the true bootstrap population mean $\mu_i^{(b)}$ for $i = 1,...,D$.
		\item[(e)] For each bootstrap population, select a bootstrap sample that matches the original sample size $n_i$ and  estimate $\hat{f}^{(b)}()$ and $\hat{v
		\nu}_i^{(b)}$. Obtain estimates for the mean $\hat{\mu}_i^{(b)}$.
	\end{enumerate}
	\item Utilizing the $B$ bootstrap samples, the MSE estimator is computed as follows:
	$$\widehat{MSE}_i = \frac{1}{B} \sum_{b=1}^{B} \biggl(\mu_i^{(b)}-\hat{{\mu}}^{(b)}_{i}\biggr)^2.$$
\end{enumerate}

\newpage
\subsection{Design-based simulation}

This section assesses the efficacy of the proposed small area estimation methods by employing a design-based simulation, using the same dataset and structure as the real-data application discussed in Section 5. The simulation was conducted with $500$ independent pseudo-survey samples drawn from the fixed population census dataset of Guerrero. Each pseudo-sample mirrors the characteristics of the original survey, maintaining the same number of in-sample municipalities. This ensures that each of the $500$ pseudo-survey samples has consistent structure and overall sample size. The true values are defined as the area-level means derived from the original census data. The estimation methods evaluated - EBPP, GMERF, and MERF - are applied as described in Section 5, using the same fixed working model for EBPP throughout the simulation.

Figures  \ref{fig:design_bias_fig} and \ref{fig:design_rmse_fig} display the results in terms of bias and RMSE for the estimated area-level means in Guerrero. A direct comparison between GMERF and MERF reveals no substantial difference in RMSE, indicating similar performance levels for both methods across different sample domains. Notably, both GMERF and MERF demonstrate superior performance in terms of RMSE when compared to EBPP, particularly highlighting the strength of non-parametric approaches in this simulation context.

Table A.1 evaluates the performance of the proposed MSE bootstrap procedures in terms of relative bias of RMSE and relative RMSE of RMSE. In terms of RRMSE-RMSE, the adjusted non-parametric bootstrap method for the MERF appears to be the most efficient method in this design-based simulation. Regardless of the bootstrap and estimation method used, the RB-RMSE for the in-sample areas indicates an acceptable bias regarding the median and mean RB-RMSE. For out-of-sample areas, we encounter underestimation in terms of the median value and overestimation according to mean values for all three methods. 

\begin{figure}[!htb]
	\centering
	\setcounter{figure}{0}  
	\renewcommand{\thefigure}{\thesection.\arabic{figure}}  
	
	\captionsetup{justification=centering,margin=1.5cm}
	\includegraphics[width=1\linewidth]{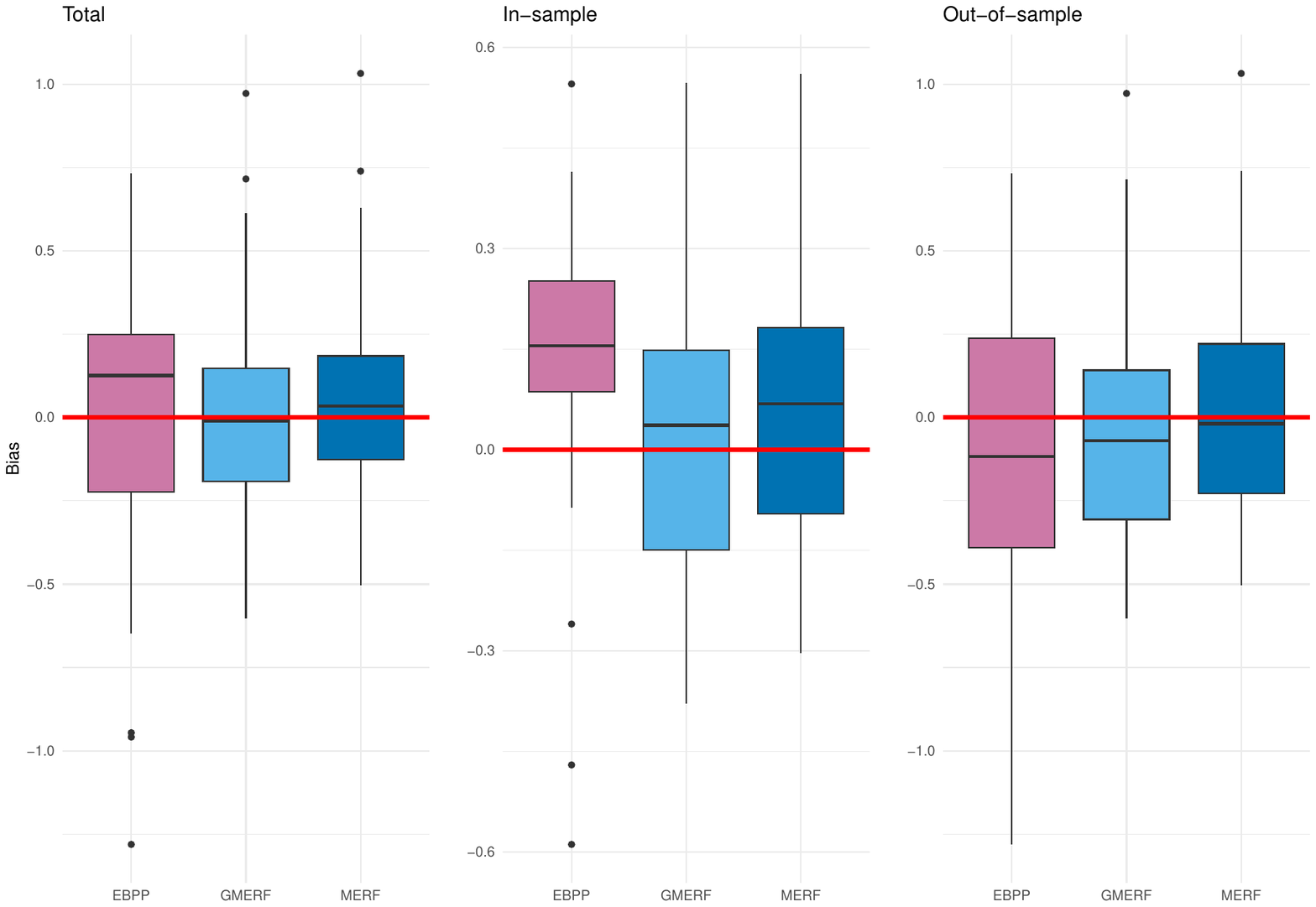}
	\caption{Bias of area-specific point estimates including details on in-
		and out-of-sample areas. Comparison of empirical bias from the design-based simulation for target variable \textit{schoolfem}.}
	\label{fig:design_bias_fig}
\end{figure}

\begin{figure}[!htb]
	\centering
	\setcounter{figure}{1}  
	\renewcommand{\thefigure}{\thesection.\arabic{figure}}  
	
	\captionsetup{justification=centering,margin=1.5cm}
	\includegraphics[width=1\linewidth]{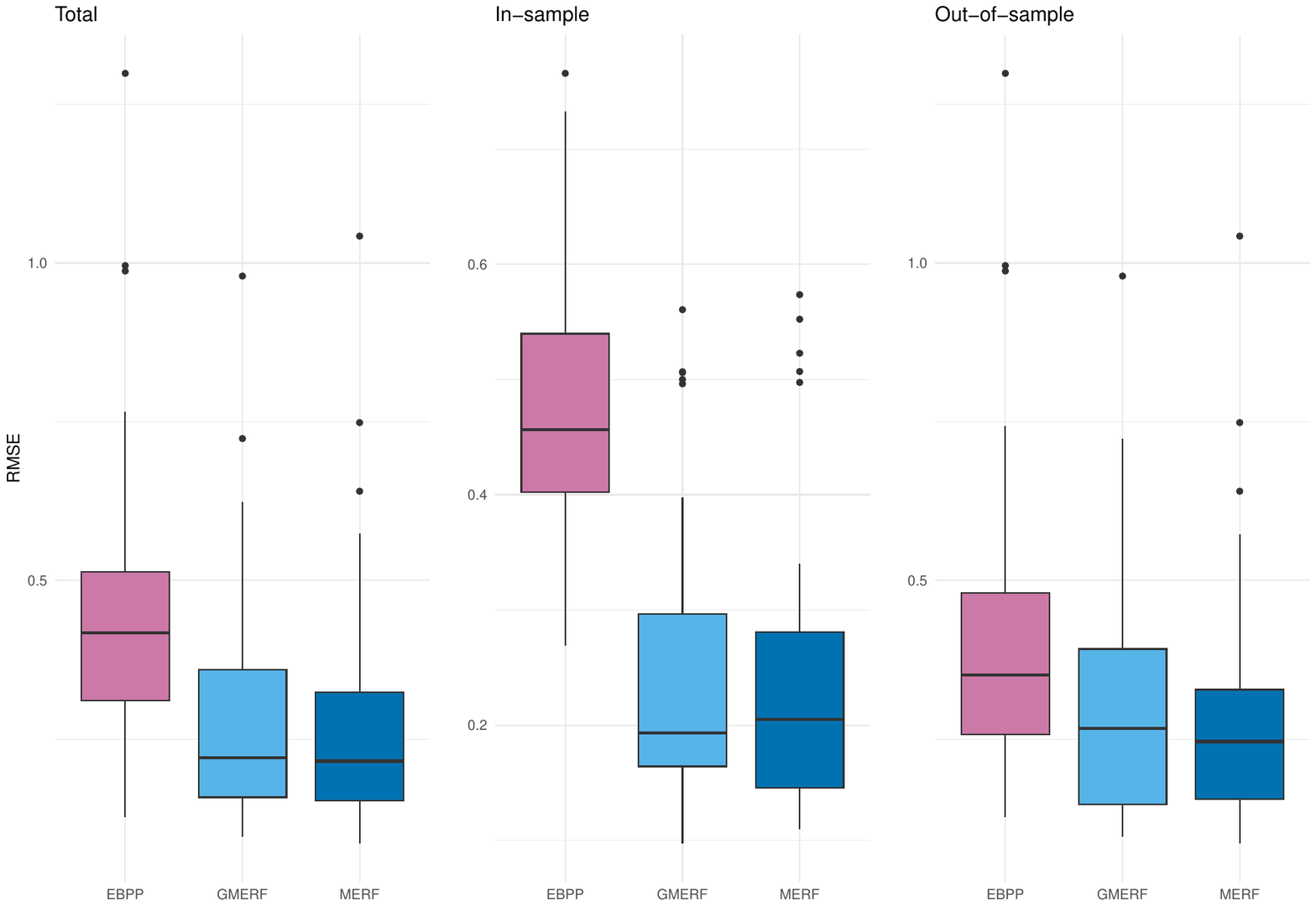}
	\caption{Performance of area-specific point estimates including details on in-
		and out-of-sample areas. Comparison of empirical RMSEs from the design-based simulation for target variable \textit{schoolfem}.}
	\label{fig:design_rmse_fig}
\end{figure}

\begin{table}[!h]
	\small
	\centering
	\setcounter{table}{0}  
	\renewcommand{\thetable}{\thesection.\arabic{table}}  
	\captionsetup{justification=centering,margin=1.5cm}
	\caption{Mean and median for relative Bias and relative RMSE of the estimated RMSE over total, in- and out-of-sample areas for MSE estimates.}
	\begin{tabular}{lcccccc}
		\\[-1.8ex]\hline
		\hline \\[-1.8ex]
		Municipalities&\multicolumn{2}{c}{Total} &\multicolumn{2}{c}{In-sample}&\multicolumn{2}{c}{Out-of-sample} \\
		\hline \\[-1.8ex]
		& Median & Mean & Median & Mean & Median & Mean \\
		\hline \\[-1.8ex]
		\multicolumn{6}{l}{RB-RMSE[\%]}\\
		\hline \\[-1.8ex]
		GMERF P &  $-5.85$ & $2.17$& $-2.95$ & $0.27$ &$-16.50$& $4.03$ \\
		GMERF NP & $-4.16 $ & $8.14$& $3.21$ & $7.19 $ & $-12.10$ & $9.07$ \\
		MERF NPC &$6.44$ & $17.59$ & $8.08$ & $13.17 $ & $-4.38$ & $21.90$\\
		\hline \\[-1.8ex]
		\multicolumn{6}{l}{RRMSE-RMSE[\%]}\\
		\hline \\[-1.8ex]			
		GMERF P & $53.63$ & $57.97$ & $45.14$ & $48.23$ & $58.93$& $67.47$\\
		GMERF NP & $51.07 $ & $58.38$ & $46.89$ & $50.61$ & $55.79$ & $65.95$\\
		MERF NPC & $42.68$ & $56.11$ & $40.84$ & $46.07$ & $45.09$ & $65.90$ \\\hline \\[-1.8ex]
	\end{tabular}
	\label{tab:design_mse_tab}
\end{table}
\end{appendices}

\end{document}